\documentclass[aps,pre,onecolumn,showpacs,showkeys,a4paper]{revtex4-1}

\input epsf  
\usepackage{graphicx} 
\usepackage{amsmath}
\usepackage{amssymb}
\usepackage{amscd}

\begin{document} 

 \title{Matrix Ansatz for the Fluctuations of the Current in the ASEP with Open Boundaries}
\author{Alexandre Lazarescu}
  \affiliation{Institut  de Physique Th\'eorique, C. E. A.  Saclay,
 91191 Gif-sur-Yvette Cedex, France}
%  \email{alexandre.lazarescu@cea.fr}
 \begin{abstract} 
 The Asymmetric Simple Exclusion Process is one of the most extensively studied models in non-equilibrium statistical mechanics. The macroscopic particle current produced in its steady state is directly related to the breaking of detailed balance, and is therefore a physical quantity of particular interest. In this paper, we build a matrix product Ansatz which allows to access the exact statistics of the fluctuations of that current for finite sizes, as well as the probabilities of configurations conditioned on the mean current. We also show how this Ansatz can be used for the periodic ASEP, and how it translates in the language of the XXZ spin chain.
  \end{abstract}

\maketitle 

\section{Introduction}

In the study of systems of many interacting particles, one of two situations might arise. If the system is isolated, or if the interaction with its environment allows it (is invariant under the reversing of time, for instance), one could observe an equilibrium state, in which the probability of any configuration is simply related to the energy of that configuration via the Gibbs-Boltzmann law. In principle, from that information, and the expression of those energies, one could calculate any equilibrium quantity that might take his fancy. Obviously, those calculations could still be extremely difficult, but the general framework provided by the Gibbs-Boltzmann law systematically solves the first part of the problem. In the other situation, where no equilibrium state can be observed, there is no such framework. The system might reach a non-equilibrium steady state, where the probabilities of the configurations do not depend on time, but there is no way to calculate those probabilities a priori, and one has to solve the whole dynamical equation that describes the evolution of the system to obtain the desired quantities.

There have been, however, many attempts to generalise the concept of Boltzmann weight to non-equilibrium systems. The best candidates to play the role of the energy in these systems are large deviation functions, which contain information on the probabilities of rare events or configurations, in the limit of some large extensive quantity (usually time or system size) \cite{Touchette, Zia, Schutzrev, Spohn, Bodineau, Bodineau1, DLSpeer}. Those large deviation functions have for instance helped uncover some very general symmetries, called `fluctuation theorems', that are verified by systems however far from equilibrium \cite{LebSpohn, Gallavotti}. The study of large deviation functions is therefore an important task to statistical physicists.

Many of those systems that display non-equilibrium steady states describe the transport of carriers (e.g. of mass, electrical charge, or thermal energy), that may interact with each-other, through some domain (a one-dimensional channel, for example), and driven by an external field in the bulk of the system, and/or unbalanced reservoirs at its boundaries. One may think, for instance, of a metallic wire conducting electrons between two masses at different potentials, or an artery conducting blood cells between two organs at different pressures. Because of those driving forces, the system exhibits a non-vanishing macroscopic current in its steady state. That current is related to the microscopic entropy production that comes from the breaking of detailed balance and time reversal invariance, and which is characteristic of non-equilibrium states (in some cases, that relation between macroscopic current and microscopic entropy production is a very simple one, as one can see in \cite{LebSpohn} and appendix \ref{GCsymm}).

One of the simplest examples of driven particle models, and one of the most extensively studied, is the Asymmetric Simple Exclusion Process (ASEP). It consists of a one-dimensional lattice, the sites of which hold particles that jump forwards and backwards stochastically. The particles interact via hard-core repulsion, so that there can be only one particle on a given site at a given time (hence `exclusion'). They jump preferentially to one direction, which accounts for the driving force in the bulk of the system (which makes it `asymmetric'). Each side of the system is connected to a particle reservoir characterised by a fixed density. The ASEP has many qualities which explain the extent to which it has been studied in the past twenty years or so \cite{MartinReview, Schutz, Varadhan, DerridaRep, KLS, Bodineau, Bodineau1, Gorissen, deGierNew, Doucot, Krug, Stinch}. First of all, it is simple in its definition, and has the mathematical property of being integrable, which makes it a good candidate for analytical calculations and exact solutions. This integrability property implies that the methods used to treat the model are usually not general and transposable (except to other integrable systems), but the actual results could give us valuable insights into the general behaviour of generic non-equilibrium systems. Moreover, the ASEP has connexions with many and various other systems, such as ribosomes travelling on a m-RNA strand \cite{PaulK, Zia, Chou, Zia2} (which is what the ASEP was originally meant to describe), random polymer models \cite{SasamSpohn}, growing interfaces \cite{KPZ, Takeuchi}, pedestrian and car traffic \cite{EvansTraffic}, quantum spin chains \cite{Sandow}, random matrices \cite{FerrariPatrick, Sasamoto, Johansson} and even pure combinatorics \cite{Corteel, Viennot}. 

In the present paper, we add our own effort to the long history of results on the steady state of the ASEP and the fluctuations of the current it exhibits, by solving the long-standing problem of obtaining the distribution of those fluctuations in the open case. Because the system is out of equilibrium, the choice of the boundary conditions matters greatly, much as for systems with long-range interactions. The statistical ensembles are not equivalent, and fixing the number of particles (as in the periodic case) or not (as in the open case) makes the results, as well as the methods applicable to their acquisition, significantly different. For instance, the full current fluctuations for the periodic TASEP (where the T stands for `totally', i.e. the particles jump in only one direction) were obtained by Bethe Ansatz in \cite{DLeb}, and the same method was used many years later for the periodic ASEP in \cite{Sylvain4}, but that method turned out to be inapplicable to the open case, precisely because of the non-conservation of the total number of particles. The Bethe Ansatz could also have led to the distribution of the steady state probabilities in the periodic case, were it not trivially flat, but the same could not be done in the open case, and another method had to be devised, namely the `matrix Ansatz' \cite{MartinReview, DEHP}, by which some recursion relations \cite{DeDoMuk} that were found in the weights of configurations of systems of different sizes are encoded in algebraic form. A first extension of that method was used to obtain the diffusion constant (second cumulant of the current) in the open ASEP \cite{DEMal}. During the last couple of years, the author and collaborators managed to generalise that method further to calculate all the cumulants in the open case, first for the TASEP \cite{LazarescuMallick}, then for the ASEP \cite{GLMV}. In both cases, the results are exact for any values of the parameters and for any finite size of the system. However, at the time, large portions of the proof were guessed rather than fully worked out, so that the results were given as a conjecture along with numerical evidence to support them. In this paper, we give the complete algebraic proof of the validity of our Ansatz, and explain how it gives us access not only to the fluctuations of the current, but also to any spatial observable conditioned on the mean current. We also show how the Ansatz extends to the periodic case, and to the spin-$\frac{1}{2}$ XXZ chain with non-diagonal boundary terms.

The layout of the paper is as follows: in section \ref{sec:Model}, we define the model, and do a quick review of some known steady state properties; in section \ref{sec:SEnsemble}, we restate the problem of finding the fluctuations of the current as that of finding the first eigenvalue of a deformed Markov matrix, and we define the `s-ensemble' as the distribution of the principal eigenstate of that same matrix; in section \ref{sec:MatAnz}, we present our perturbative matrix Ansatz, along with its proof (the technical portions of which are carried out in the appendixes), an alternative formulation which was used in \cite{GLMV}, and its equivalent for the periodic case and the XXZ chain; finally, in section \ref{sec:CurrCumul}, we give a quick overview of how the explicit calculations of the cumulants of the current were carried out using our Ansatz.

\section{Definition of the model and steady state properties}
\label{sec:Model}

\subsection{The open ASEP}
\label{sec:ModelASEP}

\begin{figure}[ht]
\begin{center}
 \includegraphics[width=0.5\textwidth]{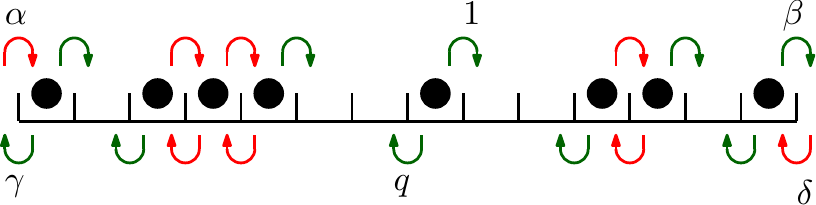}
  \caption{(colour online) Dynamical rules for the ASEP  with open boundaries. The rate
 of forward jumps has been normalised to 1. Backward jumps occur with rate
 $q < 1$. All other parameters are arbitrary. The jumps shown in green are allowed by the exclusion constraint. Those shown in red are forbidden.}
\label{fig-PASEP}
 \end{center}
 \end{figure}

The open Asymmetric Simple Exclusion Process (ASEP) in continuous time is defined as follows. Let us consider a chain of $L$ sites, numbered $1$ through $L$, each of which can hold at most one particle. The occupation of site $i$ is denoted $\tau_i$ (equal to $0$ for an empty site, and $1$ for an occupied site). Those particles can jump stochastically to the right with rate $p=1$ and to the left with rate $q<1$, provided that the receiving site be empty. In addition, each end of the chain is connected to a reservoir of particles, so that particles may enter the system at site $1$ with rate $\alpha$ or at site $L$ with rate $\delta$, and exit the system from site $1$ with rate $\gamma$ and from site $L$ with rate $\beta$ (fig. \ref{fig-PASEP}) (the rate $p$ can be set to $1$ without any loss of generality).

At any time $t$, the system can be described by the probability vector $|\!|P_{t}\rangle\!\rangle$, of which the entries $P_t({\mathcal C})$ give the probability of being in the configuration ${\mathcal C}=(\tau_i)_{1..L}$ at time $t$. This probability depends only on the initial condition $|\!|P_{0}\rangle\!\rangle$, and verifies the master equation:
\begin{equation}\label{MP}
\frac{d}{dt}|\!|P_{t}\rangle\!\rangle=M|\!|P_{t}\rangle\!\rangle
\end{equation}
solved by
\begin{equation}\label{Pt}
|\!|P_{t}\rangle\!\rangle={\rm e}^{t M}|\!|P_{0}\rangle\!\rangle
\end{equation}
where the Markow matrix $M$ is given by:
\begin{equation}
M=M_0+\sum_{i=1}^{L-1} M_{i}+M_L
\end{equation}
with $M_i$ containing the jumping rates between sites $i$ and $i+1$, and $M_0$ and $M_L$ corresponding to the couplings with the left and right reservoirs (see equ.(\ref{MMu0}) with $\mu$ set to $0$ for the explicit expression of those matrices).

Each non-diagonal entry of $M$ contains one of the aforementioned rates, and is non-zero only if the initial and final configurations differ by the jumping of exactly one particle. The diagonal entries contain (minus) the rate at which the system leaves a given configuration, which is also the sum of the rates from that configuration to any other, so that the sum of each column of $M$ is $0$ (i.e. the evolution of the system conserves probability, and $M$ is a stochastic matrix). From the Perron-Frobenius theorem, we know that the largest eigenvalue of $M$ is $0$, and that, for large times, $|\!|P_{t}\rangle\!\rangle$ converges to the corresponding right eigenvector  $|\!|P^\star\rangle\!\rangle$ (the corresponding left eigenvector is the vector with all entries equal to $1$, and will be denoted by $\langle\!\langle 1|\!|$).

\subsection{Matrix Ansatz for the steady state and phase diagram}
\label{sec:MatAnz}

The exact expression of the steady state $|\!|P^\star\rangle\!\rangle$ can be written in the form of a matrix product state (also called `matrix Ansatz') \cite{DEHP}:
\begin{equation}
P^\star({\mathcal C}) = \frac{1}{Z_L} \langle W | \prod_{i=1}^L  \left(
  \tau_i { D} + (1 - \tau_i)  { E}\right) | V  \rangle \, ,
\label{MPA}
\end{equation}
where the matrices $D$ and $E$, and the vectors $\langle W | $ and $| V  \rangle$, are defined by the following algebraic relations:
\begin{align} \label{DEHPAlgebra}
         D E -q ~E D & =   (1 -q )\left( D+E \right)  \nonumber \\  \langle W | \, ( \alpha  E - \gamma D)  & =   (1 -q ) \langle W |   \nonumber \\( \beta  D - \delta E)  \,  | V  \rangle   & =   (1 -q )  | V  \rangle
              \,\, . 
\end{align}
and $Z_L=\langle W | (D+E)^L| V  \rangle$.

Let us remark here that throughout this paper, we use the standard bra/ket notation for vectors from the space on which matrices $D$ and $E$ act (e.g. $| V  \rangle$), and the doubled bra/ket notation for vectors from the configurations space (e.g. $|\!|P^\star\rangle\!\rangle$).

As stated in (\ref{MPA}), the weight of a given configuration ${\mathcal C}=(\tau_i)_{1..L}$ in $|\!|P^\star\rangle\!\rangle$ can be written as the product of $L$ matrices $D$ or $E$, contracted between two vectors. The $i$th matrix in the product corresponds to the occupation of the $i$th site in ${\mathcal C}$: it is $D$ if $\tau_i=1$ and $E$ if $\tau_i=0$. The first algebraic relation in (\ref{DEHPAlgebra}) encodes a recursion between the steady state probabilities of systems of successive sizes. Combined with the two other relations, it allows, in principle, to compute explicit expressions of any of those probabilities, although that computation would be extremely impractical. However, it can be used, in a most elegant manner, to compute the mean values of certain observables, like the particle current passing through the system, or the local density \cite{DeDoMuk}, which are more physically relevant than the probability of a single configuration. For instance, for a system of size $L$, the stationary current $J$ is given by $J=(1-q)Z_{L-1}/Z_L$, and can be expressed in terms of certain orthogonal polynomials \cite{SasaPasep1, Corteel}. Those calculations are especially easy to carry out in the simpler case of the Totally Asymmetric Simple Exclusion Process \cite{DEHP} (TASEP, for which $q=\gamma=\delta=0$), and even more so in the case $\alpha=\beta=1$ \cite{Viennot}.

In the large size limit $L\rightarrow \infty$,  those quantities (mean current $J$ and mean density $\rho(x)$ where $x=i/L$) can be used to characterise the different phases in which the system may find itself, depending on the values of the boundary parameters. Those phases are best described in terms of the effective reservoir densities
\begin{align}
\rho_a&=1/(1+a_+)\\
\rho_b&=b_+/(1+b_+)
\end{align}
where
\begin{align}
a_{\pm}  &=  \frac{ (1-q-\alpha+\gamma)\pm
  \sqrt{(1-q-\alpha+\gamma)^2+4\alpha\gamma}}{2\alpha} \, , \\ b_{\pm}
&=  \frac{(1-q-\beta+\delta) \pm
  \sqrt{(1-q-\beta+\delta)^2+4\beta\delta}}{2\beta}\, .
\end{align}

We may note that those densities verify the relations ~$\alpha\frac{1}{\rho_a}\!-\!\gamma\frac{1}{(1-\rho_a)}\!=\!(1\!-\!q)$~ and ~$\beta\frac{1}{(1-\rho_b)}\!-\!\delta\frac{1}{\rho_b}\!=\!(1\!-\!q)$, which can be related to the second and third equations in (\ref{DEHPAlgebra}).

~~

 \begin{figure}[ht]
\begin{center}
 \includegraphics[width=0.5\textwidth]{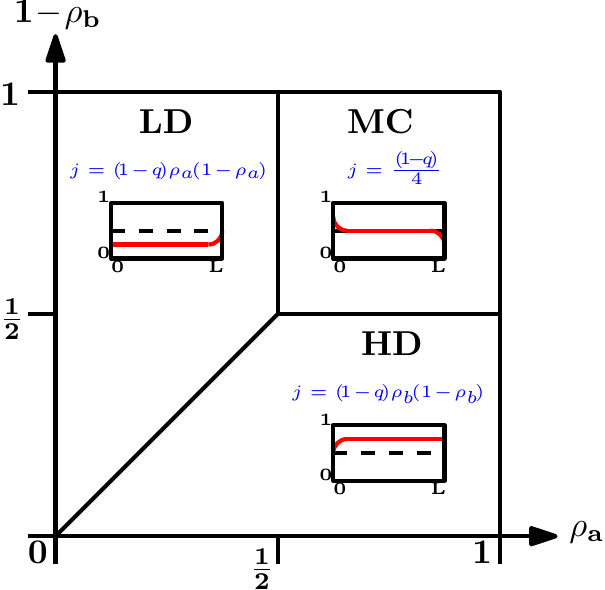}
  \caption{(colour online) Phase diagram of the open ASEP in terms of the boundary densities $\rho_a$ and $\rho_b$. The profiles in each phase show the schematic behaviour of the mean density in the system for large sizes.}
\label{fig-Diagram}
 \end{center}
 \end{figure}

The phase diagram of the open ASEP \cite{Schutz, MartinReview}, which is essentially identical to that of the TASEP up to a factor $(1-q)$ in the current, is divided in three regions (fig. \ref{fig-Diagram}):

\begin{itemize}
\item Maximal Current phase (MC): $\rho_a\!>\!1/2$ and $\rho_b\!<\!1/2$ (i.e. $a_+\!<\!1$ and $b_+\!<\!1$)~; the mean current is $J=\frac{(1-q)}{4}$~; the mean density is $1/2$ in the bulk, with an algebraic crossover to the boundary density at both ends.
\item Low Density phase (LD): $\rho_a\!<\!1/2$ and $\rho_b\!<\!1-\rho_a$ (i.e. $a_+\!>\!1$ and $b_+\!<\!a_+$)~; the mean current is $J=(1-q)\rho_a(1-\rho_a)$~; the mean density is $\rho_a$ in the bulk and at the left boundary, with an exponential crossover to $\rho_b$ at the right boundary.
\item High Density phase (HD): $\rho_a\!>\!1-\rho_b$ and $\rho_b\!>\!1/2$ (i.e. $b_+\!>\!1$ and $a_+\!<\!b_+$)~; the mean current is $J=(1-q)\rho_b(1-\rho_b)$~; the mean density is $\rho_b$ in the bulk and at the right boundary, with an exponential crossover to $\rho_a$ at the left boundary.
\end{itemize}

In order to access the fluctuations of the current, and not only its mean value, we need to find the steady state probabilities as a function of the time-integrated current as well as of the configuration. That is what we propose to do in this paper.

\section{Current-counting Markov matrix and the s-ensemble:}
\label{sec:SEnsemble}

Suppose that we want to keep track of the number of particles that jump over the bond that links sites $i$ and $i\!+\!1$ through the evolution of the system, starting from some initial state $|\!|P_{0}\rangle\!\rangle$. One easy way to do this is to multiply the off-diagonal entries of $M_i$ by a fugacity ${\rm e}^{\pm \mu_i}$, so that every time those jumping rates are used in the evolution of the system, the weight of the corresponding history gets multiplied by ${\rm e}^{\mu_i}$ if the jump was made forwards, or ${\rm e}^{-\mu_i}$ if it was made backwards. After a time $t$, the weight of any history ${\mathcal C}(t)$ carries an extra weight ${\rm e}^{J({\mathcal C}(t))\mu_i}$, where $J({\mathcal C}(t))$ is the (algebraic) number of particles that jumped from site $i$ to site $i\!+\!1$, which is precisely the time-integrated current that went through that bond. One can then access the moments of that current simply by taking derivatives with respect to $\mu_i$.

In general, one can do the same with any and all of the bonds, using different fugacities. The corresponding Markov matrix can be written as \cite{LebSpohn, DLeb2, DonskVar}:
\begin{equation}
M_{\{\!\mu_i\!\}}=M_0(\mu_0)+\sum_{i=1}^{L-1} M_{i}(\mu_i)+M_L(\mu_l)
\end{equation}
where
\begin{equation}\label{MMu}
M_0(\mu_0)=\begin{bmatrix} -\alpha & \gamma{\rm e}^{-\mu_0} \\ \alpha{\rm e}^{\mu_0} & -\gamma  \end{bmatrix}~,~ M_{i}(\mu_i)=\begin{bmatrix} 0 & 0 & 0 & 0 \\ 0 & -q & {\rm e}^{\mu_i} & 0 \\ 0 & q{\rm e}^{-\mu_i}& -1 & 0 \\ 0 & 0 & 0 & 0 \end{bmatrix}~,~M_L(\mu_l)=\begin{bmatrix} -\delta & \beta{\rm e}^{\mu_L} \\  \delta{\rm e}^{-\mu_L} & -\beta  \end{bmatrix}
\end{equation}
(it is implied that $M_0$ acts as written on site $0$ in the basis $\{0,1\}$ and as the identity on the other sites, and the same goes for $M_L$ on site $L$; similarly, $M_i$ is expressed by its action on sites $i$ and $i\!+\!1$ in the basis $\{00,01,10,11\}$ and acts as the identity on the rest of the system).

By using this deformed Markov matrix in the time evolution of $|\!|P_{t}\rangle\!\rangle$ instead of the usual one, one gets
\begin{equation}
|\!|P_{t}(\!\{\mu_i\}\!)\rangle\!\rangle={\rm e}^{t M_{\{\!\mu_i\!\}}}|\!|P_{0}\rangle\!\rangle
\end{equation}
which is the Laplace transform of the joint probabilities of the configurations and the time-integrated currents, with respect to these currents, at time $t$. Taking a $k$th order derivative in any of the $\mu_i$s and then projecting this vector onto $\langle\!\langle 1|\!|$ yields the $k$th moment of the corresponding current, up to a normalisation. 

In the long time limit, the matrix ${\rm e}^{t M_{\{\!\mu_i\!\}}}$ converges to the projector onto its principal eigenvector $|\!|P_{\{\!\mu_i\!\}}\rangle\!\rangle$, with the eigenvalue ${\rm e}^{t E(\!\{\!\mu_i\!\}\!)}$ where $E(\!\{\!\mu_i\!\}\!)$ is the largest eigenvalue of $M_{\{\!\mu_i\!\}}$, so that 
\begin{equation}
\langle\!\langle 1|\!|P_{t}(\!\{\mu_i\}\!)\rangle\!\rangle\sim {\rm e}^{t E(\!\{\!\mu_i\!\}\!)}
\end{equation}
and $E_{\{\!\mu_i\!\}}$ is therefore identified as the generating function of the cumulants of the instantaneous currents $J_i/t$, which is the Laplace transform of the large deviation function of those same currents, as explained by the Donsker-Varadhan theory of temporal large deviations \cite{Touchette, DonskVar}. That is the main quantity that we want to calculate. Our problem thus reduces to that of finding the largest eigenvalue of the current-counting matrix $M_{\{\!\mu_i\!\}}$. The corresponding eigenvector $|\!|P_{\{\!\mu_i\!\}}\rangle\!\rangle$ also holds important information, as we will see below.

~~

Let us first make things a little simpler by noticing that one can go from any set $\{\!\mu_i\!\}$ to any other set $\{\!\mu'_i\!\}$ by a matrix similarity, as long as $\sum_{i=0}^{L}\mu_i\!=\!\sum_{i=0}^{L}\mu'_i=\mu$ (see Appendix A for a detailed derivation of that statement, and \cite{LebSpohn}). That means that the eigenvalues of $M_{\{\!\mu_i\!\}}$ only depend on $\mu$, regardless of how the fugacities are distributed: the currents through each of the bonds are all exactly equivalent. In particular, there is a set for which $\mu=\lambda \log{\biggl(\frac{\gamma q^{L-1}\delta}{\alpha p^{L-1}\beta}\biggr)}$ where $\lambda$ is the quantity conjugate to the entropy production in the system (see appendix \ref{GCsymm}). This is an easy way to prove the Gallavotti-Cohen symmetry for the current in this system, and it shows that the current and its fluctuations are simply proportional to the entropy production. That entropy production being non-zero is the defining characteristic of a non-equilibrium system.

All this being said, we can now consider, without any loss of generality, the case where only the first bond (between the leftmost reservoir and the first site) is marked: $\mu_0=\mu~,~\mu_{i\neq 0}=0$, so that the individual jump matrices we will work with are given by
\begin{equation}\label{MMu0}
M_0(\mu)=\begin{bmatrix} -\alpha & \gamma{\rm e}^{-\mu} \\ \alpha{\rm e}^{\mu} & -\gamma  \end{bmatrix}~,~ M_{i}=\begin{bmatrix} 0 & 0 & 0 & 0 \\ 0 & -q & 1 & 0 \\ 0 & q& -1 & 0 \\ 0 & 0 & 0 & 0 \end{bmatrix}~,~M_L=\begin{bmatrix} -\delta & \beta \\  \delta & -\beta  \end{bmatrix}
\end{equation} 
and that
\begin{equation}
|\!|P_{t}(\!\mu\!)\rangle\!\rangle={\rm e}^{t M_\mu}|\!|P_{0}\rangle\!\rangle\sim {\rm e}^{t E(\mu)}|\!|P_\mu\rangle\!\rangle
\end{equation}
with
\begin{equation}
E(\mu)=\sum_{k=1}^{\infty}E_k\frac{\mu^k}{k!}
\end{equation}
the exponential generating series of the cumulants of the current.

The large deviation function for the instantaneous current, $G(j)\sim -\frac{1}{t}\log[{\rm P}(J/t=j)]$, is related to $E(\mu)$ by
\begin{equation}
G(j)=\mu j-E(\mu)~~,~~\frac{d}{d\mu} E(\mu)=j~.
\end{equation}
Moreover, one can show \cite{DSimon} that the principal right eigenvector $|\!|P_\mu\rangle\!\rangle$ of $M_\mu$, and its principal left eigenvector$\langle\!\langle \tilde{P}_\mu|\!|$, hold (respectively) the probabilities of observing configurations coming from the steady state, and the probabilities of having come from configurations while in the steady state, all conditioned on having observed a mean current $j=\frac{d}{d\mu} E(\mu)$:
\begin{align}
P_\mu({\mathcal C})&={\rm P}\Bigl(\{{\mathcal C},t\}\Big|\{P^\star,t\!=\!-\infty\}~\&~j\!=\!\frac{d}{d\mu} E(\mu)\Bigr)\\
\tilde{P}_\mu({\mathcal C})&={\rm P}\Bigl(\{{\mathcal C},t\}\Big|\{P^\star,t\!=\!+\infty\}~\&~j\!=\!\frac{d}{d\mu} E(\mu)\Bigr)
\end{align}
so that the product of the two is the probability of observing a configuration at any time, conditioned on the mean current (up to a normalisation):
\begin{equation}
P_\mu({\mathcal C})\tilde{P}_\mu({\mathcal C})={\rm P}\Bigl({\mathcal C}~\Big|~j\!=\!\frac{d}{d\mu} E(\mu)\Bigr)
\end{equation}

The ensemble defined by these probabilities, with $\mu$ as a parameter, is sometimes called the `s-ensemble' \cite{SEnsemble} (the reason being that $\mu$ is often noted $s$), and contains all the information needed to build the joint large deviation functions of the current and any spatial observables (i.e. depending only on ${\mathcal C}$). We will now construct a matrix ansatz that holds the exact expressions of those probabilities and of $E(\mu)$, as series in $\mu$, up to arbitrary orders.

\section{Perturbative matrix Ansatz for the s-ensemble:}
\label{sec:MatAnz}

In this section, we will show that we can define two transfer matrices $T_\mu$ and $U_\mu$, such that:
\begin{equation}\label{MUT}
\bigl[M_\mu,U_\mu T_\mu\bigr]=0
\end{equation}
\begin{equation}\label{T0}
T_0=[1]_{{\mathcal C},{\mathcal C'}}= |\!| 1\rangle\!\rangle \langle\!\langle 1|\!|
\end{equation}
\begin{equation} \label{U0}
U_0  |\!| 1\rangle\!\rangle= |\!|P^\star\rangle\!\rangle
\end{equation}
where the weights of $T_\mu$ and $U_\mu$ are expressed as products of matrices between two vectors, much as in (\ref{MPA}). Moreover, equ. (\ref{MUT}) for $\mu=0$ allows to recover that original Matrix Ansatz, as we will show below.

Those three relations will be used to prove that the transfer matrix $U_\mu T_\mu$ is almost a projector onto the leading eigenstate of $M_\mu$, and that when applied repeatedly, the precision in orders of $\mu$ of the projection increases. In other terms:

\begin{equation}\label{UTmu}
(U_\mu T_\mu)^k\sim |\!|P_\mu\rangle\!\rangle \langle\!\langle \tilde{P}_\mu|\!|+{\mathcal O}\left(\mu^{k}\right)
\end{equation}
up to a multiplicative constant of order $1$.

We will then use that relation to show the main results of this paper, namely that
 \begin{align}
 |\!| P_{\mu}\rangle\!\rangle&=\frac{1}{Z_L^{(k)}} (U_\mu T_\mu)^k    |\!| P^\star\rangle\!\rangle+ {\mathcal O}\left(\mu^{k+1}\right)
\label{MPA3}      \\
 \langle\!\langle \tilde{P}_{\mu} |\!|& =\frac{1}{Z_L^{(k)}} \langle\!\langle 1|\!| (U_\mu T_\mu)^k+ {\mathcal O}\left(\mu^{k+1}\right)
\label{MPA4}      
\end{align}
where $Z_L^{(k)}=\langle\!\langle 1|\!| (U_\mu T_\mu)^k|\!| P^\star\rangle\!\rangle$,
and that
 \begin{equation}\label{Emuk}
E(\mu)=\frac{ \langle\!\langle 1|\!| M_\mu (U_\mu T_\mu)^k  |\!|P^\star\rangle\!\rangle }{ \langle\!\langle 1|\!| (U_\mu T_\mu)^k  |\!|P^\star\rangle\!\rangle}+ {\mathcal O}\left(\mu^{k+2}\right).
\end{equation}

Those results hold for any integer $k$, so that, in essence, we have complete exact expressions for $ |\!| P_{\mu}\rangle\!\rangle$, $\langle\!\langle \tilde{P}_{\mu} |\!|$ and $E(\mu)$, expanded as infinite series in $\mu$.

\subsection{Definitions and commutation relations:}
\label{sec:MatComm}

Let us consider two matrices $d=D-1$ and $e=E-1$ where $D$ and $E$ are defined as in (\ref{DEHPAlgebra}), and a matrix $A_\mu$, such that:
\begin{align}
            de -q~e d &=  (1 -q ) \nonumber \\ 
            e A_\mu  & = {\rm e}^\mu ~ A_\mu e
             \nonumber   \\ A_\mu d  & = {\rm e}^\mu ~ d A_\mu \,\, . 
 \label{dehpAlgebra}
\end{align}

The first of these relations, which is a simple consequence of (\ref{DEHPAlgebra}), defines the algebra of a q-deformed harmonic oscillator \cite{QGroups}, of which $e$ is the creation operator, and $d$ the annihilation operator.

Let us also define two more boundary vectors $ \langle \tilde{W} |$ and $|\tilde{V}\rangle$ by:
 \begin{align} 
     [ \beta ( 1  - d)  
   - \delta ( 1  -  e ) ] \,  | \tilde{V}   \rangle  &= 0   \nonumber \\
       \langle  \tilde{W} | [  \alpha(1 - e )
  - \gamma (1 -  d )] &= 0 
   \label{VWT}
 \end{align} 
and let us recall that
\begin{align} 
     [ \beta ( 1  + d)  
   - \delta ( 1  +  e ) ] \,  |  V  \rangle  &= (1-q)|  V  \rangle  \nonumber \\
       \langle  W | [  \alpha(1 + e )
  - \gamma (1 +  d )] &= (1-q)  \langle  W |
   \label{VW}
 \end{align} 

By writing
\begin{equation}
X_{0,0}=X_{1,1}=1~,~X_{1,0}=d~,~X_{0,1}=e
\end{equation}
we can finally define the weights of $T_\mu$ and $U_\mu$ between configurations ${\mathcal C'}=(\tau'_i)_{1..L}$ and ${\mathcal C}=(\tau_i)_{1..L}$:
 \begin{equation}   \label{Umu}
U_\mu({\mathcal C},{\mathcal C'})=\frac{1}{Z_L}  \langle W| A_\mu \prod_{i=1}^{L}X_{\tau_i,\tau_i'}  | V \rangle
\end{equation}
with $Z_L= \langle W|\bigl(2+d+e\bigr)^L | V \rangle$, and
 \begin{equation}   \label{Tmu}
T_\mu({\mathcal C},{\mathcal C'})=  \langle  \tilde{W}| A_\mu \prod_{i=1}^{L}X_{\tau_i,\tau_i'}  |  \tilde{V} \rangle .
\end{equation}

Those weight are entirely determined by the algebra defined above. Specifically, one can obtain the weights of $U_\mu$ buy using (\ref{dehpAlgebra}) and (\ref{VW}), and those of $T_\mu$ buy using (\ref{dehpAlgebra}) and (\ref{VWT}).

We may note that the matrix $A_\mu$ is set between the left boundary vector and the first matrix because it is the bond between the left reservoir and the first site that is marked. For a general set of weights $\{\mu_i\}$, we would have to add matrices $A_{\mu_i}$ between $X_{\tau_i,\tau_i'}$ and $X_{\tau_{i+1},\tau_{i'+1}}$ in both of the products above, so that:

 \begin{align}   \label{Umui}
U_{\{\mu_i\}}({\mathcal C},{\mathcal C'})&=\frac{1}{Z_L}  \langle W| A_{\mu_0} \prod_{i=1}^{L}X_{\tau_i,\tau_i'} A_{\mu_i} | V \rangle\\
T_{\{\mu_i\}}({\mathcal C},{\mathcal C'})&=  \langle  \tilde{W}| A_{\mu_0} \prod_{i=1}^{L}X_{\tau_i,\tau_i'} A_{\mu_i} |  \tilde{V} \rangle .
\end{align}

~~

In appendix \ref{Commut1}, we derive equation (\ref{MUT}), using a method closely related to the matrix Ansatz for multispecies ASEP on a ring \cite{PEM}, and which makes use of the so-called `hat matrices'. That equation is the main point to our Ansatz: the transfer matrix $U_\mu T_\mu$ that we built has the same eigenvectors as $M_\mu$, so that we can try to extract the information we need from it instead of $M_\mu$.

For $\mu=0$, one particular solution to (\ref{dehpAlgebra}) and (\ref{VWT}) is $d=e=A_0=1$, so that for any ${\mathcal C'}$ and ${\mathcal C}$, we have $T_0({\mathcal C},{\mathcal C'})=\langle  \tilde{W}| \tilde{V} \rangle$ which we can set to $1$. This proves (\ref{T0}). Furthermore, projecting $U_0$ onto $|\!| 1\rangle\!\rangle$ means summing over all configurations ${\mathcal C'}$ in (\ref{Umu}), so that
\begin{equation}
\langle\!\langle {\mathcal C}|\!|U_0  |\!| 1\rangle\!\rangle=\sum_{{\mathcal C'}}U_\mu({\mathcal C},{\mathcal C'})= \langle W| A_0 \prod_{i=1}^{L}(X_{\tau_i,0}+X_{\tau_i,1})  | V \rangle
\end{equation}

We can set $A_0$ to $1$, and remark that for $\tau_i=0$, we have $(X_{\tau_i,0}+X_{\tau_i,1})=1+e=E$ and that for $\tau_i=1$, we have $(X_{\tau_i,0}+X_{\tau_i,1})=d+1=D$, so that this expression is exactly that of $P^\star({\mathcal C})$ as given in (\ref{MPA}), which proves (\ref{U0}).

Using relations (\ref{MUT}), (\ref{T0}) and (\ref{U0}) together, at $\mu=0$, we get
\begin{equation}
[M,U_0 T_0]=0=\Bigl(M |\!|P^\star\rangle\!\rangle \langle\!\langle 1|\!|\Bigr)-\Bigl(|\!|P^\star\rangle\!\rangle \langle\!\langle 1|\!|M\Bigr)~.
\end{equation}
Since we know that $\langle\!\langle 1|\!|M=0$ (because $M$ is a stochastic matrix), this implies that $M |\!|P^\star\rangle\!\rangle=0$, which yields the original matrix Ansatz (\ref{MPA}) \cite{DEHP}.

This alternative proof of (\ref{MPA}) relies on the fact that the transfer matrix $U_\mu T_\mu$ is a projector in the limit $\mu\rightarrow 0$. It would be interesting to determine whether for other situations with matrix product states, one can generically find a transfer matrix that commutes with a deformation of the dynamics of the system and is a projector in the non-deformed limit. One could for instance look at the ASEP in discrete time with different versions of the update \cite{DiscTime}, or at the multispecies ASEP on a ring \cite{PEM}.

\subsection{Validation of the perturbative matrix Ansatz:}
\label{sec:PertAns}

To prove (\ref{MPA3}) and (\ref{Emuk}), we use the relations derived above. Since, for $\mu=0$, the matrix $U_0 T_0$ is the projector onto the principal eigenspace of $M$, one can write, for an infinitesimal $\mu$:
\begin{equation}
U_\mu T_\mu=\Lambda_\mu |\!|P_\mu\rangle\!\rangle \langle\!\langle \tilde{P}_\mu|\!|+r_\mu
\end{equation}
where $\Lambda_\mu\sim 1+ {\mathcal O}\left(\mu\right)$ is the largest eigenvalue of $U_\mu T_\mu$, and $r_\mu\sim {\mathcal O}\left(\mu\right)$ is the part of $U_\mu T_\mu$ that is orthogonal to its principal eigenspace, and has eigenvalues of order $\mu$. In other words, $U_\mu T_\mu$ is almost a projector, with an error $r_\mu$ of order $\mu$.

Since $r_\mu  |\!|P_\mu\rangle\!\rangle=0$ and $\langle\!\langle \tilde{P}_\mu|\!|r_\mu =0$, one has that:
\begin{equation}
(U_\mu T_\mu)^k=\Lambda_\mu^k |\!|P_\mu\rangle\!\rangle \langle\!\langle \tilde{P}_\mu|\!|+r_\mu^k
\end{equation}
so that the difference from the projector onto $|\!|P_\mu\rangle\!\rangle \langle\!\langle \tilde{P}_\mu|\!|$ is now $r_\mu^k\sim  {\mathcal O}\left(\mu^k\right)$.

Let us now remark that the parts of $|\!|P^\star\rangle\!\rangle$ and of $\langle\!\langle 1 |\!|$ which are not in the kernel of $r_\mu$ are of order $\mu$, so that both $r_\mu^k|\!|P^\star\rangle\!\rangle$ and $\langle\!\langle 1 |\!|r_\mu^k$ are of order $\mu^{k+1}$. It follows that $(U_\mu T_\mu)^k|\!|P^\star\rangle\!\rangle$ is proportional to $|\!|P_\mu\rangle\!\rangle$ with an error of order $\mu^{k+1}$  (and the same goes for $\langle\!\langle \tilde{P}_\mu|\!|$), which proves (\ref{MPA3}).

Equation (\ref{Emuk}) is then proven by simply applying $M_\mu$ to $(U_\mu T_\mu)^k|\!|P^\star\rangle\!\rangle$:
\begin{align}
\langle\!\langle 1|\!| M_\mu (U_\mu T_\mu)^k|\!|P^\star\rangle\!\rangle &=E(\mu)\Lambda_\mu^k \langle\!\langle 1|\!|P_\mu\rangle\!\rangle\langle\!\langle \tilde{P}_\mu|\!|P^\star\rangle\!\rangle+\langle\!\langle 1|\!| M_\mu r_\mu^k|\!|P^\star\rangle\!\rangle\\
\langle\!\langle 1 |\!|(U_\mu T_\mu)^k|\!|P^\star\rangle\!\rangle &=\Lambda_\mu^k\langle\!\langle 1|\!|P_\mu\rangle\!\rangle\langle\!\langle \tilde{P}_\mu|\!|P^\star\rangle\!\rangle+\langle\!\langle 1|\!| r_\mu^k|\!|P^\star\rangle\!\rangle
\end{align}
where $\langle\!\langle 1|\!| M_\mu r_\mu^k|\!|P^\star\rangle\!\rangle$ is of order $\mu^{k+2}$ because $\langle\!\langle 1|\!| M_\mu$ is of order $\mu$ and $r_\mu^k|\!|P^\star\rangle\!\rangle$ is of order $\mu^{k+1}$, and $\langle\!\langle 1 |\!|r_\mu^k|\!|P^\star\rangle\!\rangle$ is of order $\mu^{k+2}$ for the reason given above. The ratio of those two  equations is therefore equal to $E(\mu)$ up to order $\mu^{k+2}$.

\subsection{Formulation as a matrix product:}
\label{sec:MatProd}

The formulation we gave here of the perturbative matrix Ansatz in terms of transfer matrices is quite different from that which was given in \cite{GLMV}. They are of course equivalent, which is what we will show in this section.

\begin{figure}[ht]
\begin{center}
 \includegraphics[width=0.6\textwidth]{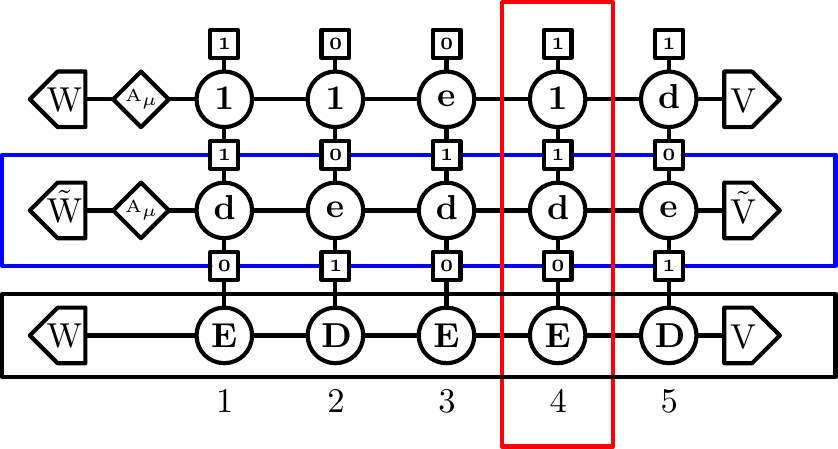}
  \caption{(colour online) One of the tensor networks that compose $U_\mu T_\mu  |\!|P^\star\rangle\!\rangle$ when expanded in terms of the intermediate configurations at each step of the product. Lines represent the transfer matrices $U_\mu$ and $T_\mu$, whereas columns represent $E_k$ or $D_k$. See the detailed explanation below.}
\label{fig-Tensors}
 \end{center}
 \end{figure}

The main point that needs to be made here is that, unlike the Markov matrix, which is a sum of elementary matrices, the transfer matrices $U_\mu$ and $T_\mu$ are products of the elementary matrices $X_{\tau_{i+1},\tau_{i'+1}}$, so that a product of those transfer matrices can be seen as a tensor network, the tensors being of order four: $\bigl(X_{\tau_{i+1},\tau_{i'+1}}\bigr)_{i,j}$, where $i$ and $j$ are the internal indices of $X$ (fig. \ref{fig-Tensors}).

A consequence of this is that the object $(U_\mu T_\mu)^k  |\!|P^\star\rangle\!\rangle$ can be written in terms of the columns of this tensor network instead of the lines. Let us therefore define, by recursion (and denote the product between successive rows by a tensor product $\otimes$):
 \begin{align}\label{DEk}
       D_{k+1} &=  (1 \otimes 1 + d \otimes e ) \otimes  D_k + ( 1 \otimes d   +  d \otimes 1)  \otimes  E_k   \nonumber  \\
       E_{k+1}  &=  (1 \otimes 1 +  e   \otimes d )  \otimes  E_k + (e  \otimes 1 + 1 \otimes  e )  \otimes   D_k  \nonumber \\
	A_\mu^{(k+1)}&=A_\mu\otimes A_\mu\otimes A_\mu^{(k)}
\end{align} 
with $D_0=D$, $E_0=E$ and $A_\mu^{(0)}=1$, and
\begin{align} 
   | V_{k+1}  \rangle  &=   | V \rangle\otimes  |\tilde{V} \rangle \otimes | V_{k}  \rangle  \\
  \langle W_{k+1} |    &= \langle W| \otimes\langle  \tilde{W} |\otimes \langle W_{k} |  \,,
\end{align} 
with $ | V_0  \rangle= | V \rangle$ and $ \langle W_0 |= \langle W |$.

In this formalism, equation (\ref{MPA3}) becomes:
 \begin{equation}
\langle\!\langle{\mathcal C} |\!| P_{\mu}\rangle\!\rangle =     \frac{1}{Z_L^{(k)}} \langle W_k |A_\mu^{(k)} \prod_{i=1}^L  \left(
  \tau_i {D_k} + (1 - \tau_i)  {E_k}\right) | V_k  \rangle 
+ {\mathcal O}\left(\mu^{k+1}\right)
\label{MPA2}      
\end{equation}

~~

To give a simple explicit example, in (fig \ref{fig-Tensors}) is shown one of the tensor networks which compose $U_\mu T_\mu  |\!|P^\star\rangle\!\rangle$ for $L=5$, namely $U_\mu(\mathcal{C},\mathcal{C'})T_\mu(\mathcal{C'},\mathcal{C''})P^\star(\mathcal{C''})$, with $\mathcal{C}=\{1,0,0,1,1\}$, $\mathcal{C'}=\{1,0,1,1,0\}$ and $\mathcal{C''}=\{0,1,0,0,1\}$. The blue rectangle corresponds to $T_\mu(\mathcal{C'},\mathcal{C''})$, the black one to $P^\star(\mathcal{C''})$, and the red rectangle is one of the elements in $D_1$, namely $X_{1,1}\otimes X_{1,0}\otimes E$. Summing over the second and third indices in any column gives $E_1$ or $D_1$, depending on whether the first (upper) index is $0$ or $1$.

~~

While the transfer matrix formulation (\ref{MPA3}) is better suited to the algebraic proof of the Ansatz, this matrix product formulation (\ref{MPA2}) is useful for doing explicit calculations, like those of the cumulants of the current (see section \ref{sec:CurrCumul}). Let us also note that the matrices $E_k$ and $D_k$ are related to the ones used in the matrix Ansatz solution of the multispecies periodic ASEP \cite{PEM}, although for the moment we have no understanding of why that is the case.

\subsection{Periodic case and XXZ spin chain:}
\label{sec:PerSymm}

The same Ansatz can be applied to the periodic case with just one alteration: instead of projecting the matrix products between boundary vectors, one has to take a trace. Moreover, only one transfer matrix $T_\mu^{per}$ needs to be defined. This can be written as
 \begin{equation}   \label{TmuPer}
T_\mu^{per}({\mathcal C},{\mathcal C'})={\rm Tr}\Bigl[ A_\mu \prod_{i=1}^{L}X_{\tau_i,\tau_i'}  \Bigr]
\end{equation}
if the marked bond is the one between sites $L$ and $1$. One can then show that $[M_\mu,T_\mu^{per}]=0$ and the rest follows, by replacing $U_\mu T_\mu$ by $T_\mu^{per}$ in every equation, using the steady state $|\!| 1_N\rangle\!\rangle$ (with coefficient $1$ for all configurations with $N$ particles), and making only one tensor product per order in (\ref{DEk}). See appendix \ref{CommutPer} for all the derivations related to the periodic case.

~~

In appendix \ref{XXZ}, we show how a special choice of the parameters ${\{\!\mu_i\!\}}$ allows our Ansatz to be applied to the spin-$\frac{1}{2}$ XXZ chain with non-diagonal boundaries. We also note that the structure of our construction is strikingly similar to that used in \cite{SchutzXXZ} to solve the Lindblad equation for the XXZ chain.

\section{Calculating the cumulants of the current (a quick overview):}
\label{sec:CurrCumul}

The principal interest of the Ansatz we constructed is that it allows to calculate the expected values of some observables without having to diagonalise $M_\mu$ explicitly. The question is, then, how we can use formula (\ref{Emuk}) to obtain explicit expressions of the cumulants of the current. As one can see in \cite{LazarescuMallick, GLMV}, we did manage to make that calculation. However, a certain amount of guesswork was used, and we believe that there is a simpler and more compact way to do it than the one we used. For that reason, we will not expose here the full detail of those (rather tedious) calculations, but rather a quick overview of the principles behind it. We hope to be able to do the full calculation in an elegant and self-sufficient way in the near future, and possibly to do the same for observables other than the cumulants of the current (as we said earlier, we should have access to any spatial observable in the s-ensemble).

\subsection{Expressing of $E(\mu)$ and $\mu$ as parametric infinite series:}
\label{sec:Series}

The main point of our reasoning from here on is that we expect the solution to have the same structure as in the periodic case \cite{DLeb}, i.e. we expect to be able to write $E(\mu)$ and $\mu$ as two infinite logarithmic series in a parameter $B$ which goes to $0$ with $\mu$:
\begin{align}\label{Dk}
E(\mu)&=-\sum_{k=1}^{\infty}D_k\frac{B^k}{k}
\\
\label{Ck}
\mu&=-\sum_{k=1}^{\infty}C_k\frac{B^k}{k}
\end{align}
 From calculations using our Ansatz in the periodic TASEP, and comparing them to the results from the Bethe Ansatz \cite{DLeb}, we were able to determine that this parameter $B$ is proportional to $\frac{(1-{\rm e}^{-\mu})}{\Lambda_\mu}$, where we recall that $\Lambda_\mu$ is the largest eigenvalue of $(U_\mu T_\mu)$ (which is the one associated with $ |\!| P_\mu\rangle\!\rangle$, and goes to $1$ when $\mu$ goes to $0$).

Luckily, there is a way to write $ |\!| P_\mu\rangle\!\rangle$ as a series in $\frac{(1-{\rm e}^{-\mu})}{\Lambda_\mu}$, from which we could get the formulae we are looking for. Let us first define
\begin{equation}\label{UTtilde}
\widetilde{U_\mu T_\mu}=\frac{1}{(1-{\rm e}^{-\mu})}\Bigl(U_\mu T_\mu- |\!| P^\star\rangle\!\rangle \langle\!\langle 1|\!|\Bigr)
\end{equation}
which is finite for $\mu\rightarrow 0$. We can now write (by taking formally the limit $k\rightarrow\infty$ in (\ref{MPA3}))
\begin{equation}
|\!| P_\mu\rangle\!\rangle= \frac{U_ \mu T_\mu}{\Lambda_\mu}|\!| P_\mu\rangle\!\rangle=\frac{\Bigl(|\!| P^\star\rangle\!\rangle \langle\!\langle 1|\!|+(1-{\rm e}^{-\mu})\widetilde{U_\mu T_\mu}\Bigr)}{\Lambda_\mu}|\!| P_\mu\rangle\!\rangle
\end{equation}
or equivalently
\begin{equation}
\Lambda_\mu|\!| P_\mu\rangle\!\rangle=\Bigl(1-\frac{(1-{\rm e}^{-\mu})}{\Lambda_\mu}\widetilde{U_\mu T_\mu}\Bigr)^{-1}|\!| P^\star\rangle\!\rangle=\sum_{k=0}^{\infty}(\widetilde{U_\mu T_\mu})^k|\!| P^\star\rangle\!\rangle \Biggl(\frac{(1-{\rm e}^{-\mu})}{\Lambda_\mu}\Biggr)^k
\end{equation}
which is a well defined series in $\frac{(1-{\rm e}^{-\mu})}{\Lambda_\mu}$.

From this, we get
\begin{equation}\label{DkTU}
E(\mu)=\langle\!\langle 1|\!|M_\mu|\!| P_\mu\rangle\!\rangle=\sum_{k=0}^{\infty}\frac{\langle\!\langle 1|\!|M_\mu(\widetilde{U_\mu T_\mu})^k|\!| P^\star\rangle\!\rangle}{(1-{\rm e}^{-\mu})}\Biggl(\frac{(1-{\rm e}^{-\mu})}{\Lambda_\mu}\Biggr)^{k+1}
\end{equation}
where $\frac{\langle\!\langle 1|\!|M_\mu(\widetilde{U_\mu T_\mu})^k|\!| P^\star\rangle\!\rangle}{(1-{\rm e}^{-\mu})}$ is finite for $\mu\rightarrow 0$ because $\langle\!\langle 1|\!|M_\mu\sim\mu$.

We also get, tautologically (since $\langle\!\langle 1|\!|P_\mu\rangle\!\rangle=1$),
\begin{equation}\label{CkTU}
\mu=-\log\Bigl[1-(1-{\rm e}^{-\mu})\langle\!\langle 1|\!|P_\mu\rangle\!\rangle)\Bigr]=-\log\Biggl[1-\sum_{k=0}^{\infty}\langle\!\langle 1|\!|(\widetilde{U_\mu T_\mu})^k|\!| P^\star\rangle\!\rangle \Biggl(\frac{(1-{\rm e}^{-\mu})}{\Lambda_\mu}\Biggr)^{k+1}\Biggr]
\end{equation}

which we can then expand in $\frac{(1-{\rm e}^{-\mu})}{\Lambda_\mu}$.

\subsection{Inferring the final formulae:}
\label{sec:Infer}

There is a major difference between expressions (\ref{CkTU}) and (\ref{Ck}) (or between (\ref{DkTU}) and (\ref{Dk})): the coefficients $C_k$ and $D_k$ should not depend on $\mu$, but $\frac{\langle\!\langle 1|\!|M_\mu(\widetilde{U_\mu T_\mu})^k|\!| P^\star\rangle\!\rangle}{(1-{\rm e}^{-\mu})}$, for instance, does. From this point on, the calculations become less precise. The reasoning is as follows:
\begin{itemize}
 \item we postulate that the coefficients $C_k$ and $D_k$ should be a somewhat identifiable part of, respectively,$\langle\!\langle 1|\!|(U_\mu T_\mu)^k|\!| P^\star\rangle\!\rangle$ and $\langle\!\langle 1|\!|M_\mu(U_\mu T_\mu)^k|\!| P^\star\rangle\!\rangle$;
\item we then calculate the equivalent terms for the periodic TASEP (which is the simplest case to which our Ansatz applies), namely $\langle\!\langle 1_N|\!|(T_\mu^{per})^k|\!| 1_N\rangle\!\rangle$ and $\langle\!\langle 1_N|\!|M_\mu^{per}(T_\mu^{per})^k|\!| 1_N\rangle\!\rangle$, for $k=2$, using the matrix product formalism and the q-deformed oscillator algebra (which, for $q=0$, becomes a random walk on $\mathbb{N}$);
\item we compare the result with the coefficients from \cite{DLeb}, and identify in which part of the calculation they emerge;
\item we isolate the corresponding part of $\langle\!\langle 1|\!|(U_\mu T_\mu)^k|\!| P^\star\rangle\!\rangle$ and $\langle\!\langle 1|\!|M_\mu(U_\mu T_\mu)^k|\!| P^\star\rangle\!\rangle$, and inject it in (\ref{Ck}) and (\ref{Dk});
\item we check numerically (i.e. do exact calculations on small sizes and orders of $\mu$) that our conjecture is correct.
\end{itemize}

The results of this calculation for the open TASEP can be found in \cite{LazarescuMallick}. The generalisation to the open ASEP \cite{GLMV} comes from applying the same reasoning to the periodic ASEP \cite{Sylvain4}; in that case, we also checked our results against DMRG calculations for low order cumulants ($E_2$ to $E_4$) and larger sizes of the system (up to $100$ sites).

\section{Conclusion:}
\label{sec:Concl}

In the present paper, we define and expose the algebraic proof of a matrix Ansatz which gives access to the principal eigenvalue and eigenvectors of the current-counting Markov matrix of the open ASEP, for any size and any value of the parameters. Using this Ansatz, the author and collaborators were able to obtain exact expressions for the cumulants of the current in the open TASEP \cite{LazarescuMallick} and the open ASEP \cite{GLMV}, which had been an open question for many years in the field of non-equilibrium statistical physics.

Much remains to be done on this subject, and we believe that this Ansatz still has a lot to offer. For instance, as was argued in section \ref{sec:SEnsemble}, the principal eigenvectors of $M_\mu$ hold the probabilities of observing a given configuration conditioned on the current flowing through the system; in other words, it should allow us to analyse the best profiles (in the sense of most probable) to produce a given atypical current. The question of finding the optimal path to produce a rare event is an important one, notably in the context of complex chemical reactions, and much work has been done to find algorithms that produce this optimal path \cite{Kurchan}. Obtaining an exact analytical result on that type of problem could provide valuable insight or help devise simpler algorithms.

Another situation where our method could be of use is the Symmetric Exclusion Process, for which many results are known for the large size limit, and have been obtained using a coarse-grained description of the system named `Macroscopic Fluctuation Theory' \cite{Bertini} and the related `additivity principle' \cite{Bodineau, Bodineau1}, but only a few exist for finite sizes \cite{AppDerr}. The limit $q\rightarrow 1$ cannot be taken directly in our results, but the present Ansatz can still be applied to the symmetric case with a few crucial alterations. However, we have yet to analyse that case in detail, which we intend to do in the near future.

There remains also the question of determining how specific or general the method we have applied here could be. We have shown that by defining the transfer matrix $U_\mu T_\mu$ which commutes with $M_\mu$, and then taking $\mu$ to $0$, one can retrieve the original matrix Ansatz \cite{DEHP}. It would be interesting to know whether the same procedure can be applied to other models with matrix product eigenstates, and if there is anything general or generalisable to it. We would also like to have a clear idea of the physical significance of that transfer matrix, which we used as a mere calculation tool, but might be interesting in itself.

~~

The author would like to thank K. Mallick and R. Vasseur for useful discussions and for reading the manuscript, as well as F. van Wijland and V. Pasquier for helpful comments.

\newpage

\appendix

\section{Gallavotti-Cohen symmetry for the currents}
\label{GCsymm}

In this section, we prove that two current-counting Markov matrices $M_{\{\mu_i\}}$ and $M_{\{\mu_i'\}}$ ar similar (and therefore have the same eigenvalues) as long as $\sum_{i=0}^{L}\mu_i=\sum_{i=0}^{L}\mu_i'$.

Let us consider the diagonal matrix $R_i(\nu_i)$ (with $1\leq i\leq L$) which multiplies by ${\rm e}^{\nu_i}$ all configurations for which site $i$ is occupied. The transformation $R_i(\nu_i)^{-1}M_{\{\mu_i\}}R_i(\nu_i)$ acts only on $M_{i-1}(\mu_{i-1})$ and $M_{i}(\mu_i)$, and a trivial calculation gives
\begin{equation}
R_i(\nu_i)^{-1}M_{i-1}(\mu_{i-1})R_i(\nu_i)=M_{i-1}(\mu_{i-1}-\nu_i)
\end{equation}
and
\begin{equation}
R_i(\nu_i)^{-1}M_{i}(\mu_{i})R_i(\nu_i)=M_{i}(\mu_{i}+\nu_i)
\end{equation}

One can therefore transfer any fraction $\nu$ of $\mu=\sum_{i=0}^{L}\mu_i$ from one bond to the previous or the next one. Using this, one can go from ${\{\mu_i\}}$ to ${\{\mu_i'\}}$ step by step, or simply derive the global similarity matrix $R_{\{\mu_i\}}^{\{\mu'_i\}}$ such that
\begin{equation}
M_{\{\mu_i'\}}=\Bigl(R_{\{\mu_i\}}^{\{\mu'_i\}}\Bigr)^{-1}M_{\{\mu_i\}}R_{\{\mu_i\}}^{\{\mu'_i\}}
\end{equation}
which one can easily find to be
\begin{equation}
R_{\{\mu_i\}}^{\{\mu'_i\}}=\prod_{i=1}^{L}R_i\Bigl(\sum_{j=0}^{i-1}(\mu_j-\mu_j')\Bigr)
\end{equation}

There is a particular set of weights $\{\mu_i\}$ defined by
\begin{equation}
\{\mu_0=\lambda\log{\biggl(\frac{\gamma}{\alpha}\biggr)},~~\mu_i=\lambda\log{(q)},~~\mu_L=\lambda\log{\biggl(\frac{\delta}{\beta}\biggr)} \}
\end{equation}
for which $M_{\{\mu_i\}}$ becomes:
\begin{equation}\label{MLambda}
M_0(\mu_0)=\begin{bmatrix} -\alpha & \gamma^{1-\lambda}\alpha^\lambda \\ \alpha^{1-\lambda}\gamma^\lambda & -\gamma  \end{bmatrix}~,~ M_{i}(\mu_i)=\begin{bmatrix} 0 & 0 & 0 & 0 \\ 0 & -q & q^\lambda & 0 \\ 0 & q^{1-\lambda}& -1 & 0 \\ 0 & 0 & 0 & 0 \end{bmatrix}~,~M_L(\mu_l)=\begin{bmatrix} -\delta & \beta^{1-\lambda}\delta^\lambda \\  \delta^{1-\lambda}\beta^\lambda & -\beta  \end{bmatrix}
\end{equation}
which is the deformed Markov matrix measuring the entropy production. We see immediately that
\begin{equation}
M_{1-\lambda}= ~^t\!M_\lambda
\end{equation}
which proves the Gallavotti-Cohen symmetry for the eigenvalues and between the left and right eigenvectors of $M_\lambda$ with respect to the transformation $\lambda\leftrightarrow(1\!-\!\lambda)$.

Considering that $\mu=\lambda \log{\biggl(\frac{\gamma \delta}{\alpha\beta}q^{L-1}\biggr)}$, we also obtain the Gallavotti-Cohen symmetry related to the current, namely
\begin{equation}
E(\mu)=E\Biggl( \log{\biggl(\frac{\gamma \delta}{\alpha\beta}q^{L-1}\biggr)}-\mu\Biggr)
\end{equation}
which is also valid for the other eigenvalues of $M_\mu$, and the corresponding relations between the right and left eigenvectors.

The equilibrium case, where $\frac{\gamma \delta}{\alpha\beta}q^{L-1}=1$, is somewhat pathological: there is no entropy production, and all the odd cumulants of the current vanish. However, by expanding all these relations to second order in the small parameter $\log{\biggl(\frac{\gamma \delta}{\alpha\beta}q^{L-1}\biggr)}$, one can recover the fluctuation-dissipation theorem.

For more results on the Gallavotti-Cohen relation, one may refer to \cite{LebSpohn}.

\section{Derivation of equation (\ref{MUT}) for the open ASEP}
\label{Commut1}

The derivation of the commutation relation (\ref{MUT}) can be carried out in two steps: first, we express the commutator of $M_i$ (for $1\leq i\leq L$) with either $U_\mu$ or $T_\mu$ (the two results are similar) using (\ref{dehpAlgebra}), and show that for both, the sum of those commutators cancels out except for two terms, related to each of the boundaries. Secondly, we check that those boundary terms, as they appear in the commutator of $M_\mu$ with the product $U_\mu T_\mu$,
cancel out as well, using (\ref{VWT}) and (\ref{VW}).

For convenience, we will here write $U_\mu$ and $T_\mu$ as:
 \begin{align}   \label{XTUmu}
U_\mu&=\frac{1}{Z_L}  \langle W| A_\mu \prod_{i=1}^{L}X^{(i)} | V \rangle\\
T_\mu&= \langle  \tilde{W}| A_\mu \prod_{i=1}^{L}X^{(i)} |  \tilde{V} \rangle .
\end{align}
with
\begin{equation}
X^{(i)}=\begin{bmatrix} 1 & e\\ d & 1\end{bmatrix}
\end{equation}
~~

Let us therefore consider $[M_i,U_\mu]$ and $[M_i,T_\mu]$. The elementary matrix $M_i$ acts only on sites $i$ and $i+1$, so that we only have to consider the commutation with the part of the matrix products that corresponds to those sites. In both $U_\mu$ and $T_\mu$, that part is $X^{(i)}X^{(i+1)}$. Let us write its components in the same basis as $M_i$ in (\ref{MMu0}), and recall the expression of $M_i$:
\begin{equation}\label{XXandM}
X^{(i)}X^{(i+1)}=\begin{bmatrix} 1 & e & e & ee \\ d & 1 & ed & e \\ d & de & 1 & e \\ dd & d & d & 1 \end{bmatrix}~~,~~M_{i}=\begin{bmatrix} 0 & 0 & 0 & 0 \\ 0 & -q & 1 & 0 \\ 0 & q& -1 & 0 \\ 0 & 0 & 0 & 0 \end{bmatrix}
\end{equation}

We now calculate:
\begin{align}\label{MXX}
[M_i, X^{(i)}X^{(i+1)}]&=\begin{bmatrix} 0 & 0 & 0 & 0 \\ (1-q)d & de-q & 1-q~ed & (1-q)e \\  (q-1)d & q-de & q~ed-1 & (q-1)e \\ 0 & 0 & 0 & 0 \end{bmatrix}-\begin{bmatrix} 0 & 0 & 0 & 0 \\ 0 & q(ed-1) & 1-ed & 0 \\ 0 & q(1-de) & de-1 & 0 \\ 0 & 0 & 0 & 0 \end{bmatrix}\nonumber\\
&=\begin{bmatrix} 0 & 0 & 0 & 0 \\ (1-q)d & de-q~ed & (1-q)ed & (1-q)e \\  (q-1)d & (q-1)de & q~ed-de & (q-1)e \\ 0 & 0 & 0 & 0 \end{bmatrix}\nonumber\\
&=(1-q)\begin{bmatrix} 0 & 0 & 0 & 0 \\ d & 1 & ed & e \\  -d & -de & -1 & -e \\ 0 & 0 & 0 & 0 \end{bmatrix}
\end{align}
by using the first relation in (\ref{dehpAlgebra}) to go from the second line to the third.

Let us now define the `hat' matrices $\hat{X}$ by $\hat{X}_{\tau,\tau'}=(-1)^{\tau}\frac{(1-q)}{2}X_{\tau,\tau'}$:
\begin{equation}
\hat{X}^{(i)}=\frac{(1-q)}{2}\begin{bmatrix} 1 & e\\ -d & -1\end{bmatrix}=\frac{(1-q)}{2}\begin{bmatrix} 1 & 0\\ 0 & -1\end{bmatrix}\centerdot X
\end{equation}
(where we note $\centerdot$ the product in the 2-dimensional space corresponding to the occupation number on one site).

If we replace the first or the second $X$ in $X^{(i)}X^{(i+1)}$ by $\hat{X}$, we get:
\begin{equation}\label{XXhat}
\hat{X}^{(i)}X^{(i+1)}=\frac{(1-q)}{2}\begin{bmatrix} 1 & e & e & ee \\ d & 1 & ed & e \\ -d & -de & -1 & -e \\ -dd & -d & -d & -1 \end{bmatrix}~~,~~X^{(i)}\hat{X}^{(i+1)}=\frac{(1-q)}{2}\begin{bmatrix} 1 & e & e & ee \\ -d & -1 & -ed & -e \\ d & de & 1 & e \\ -dd & -d & -d & -1 \end{bmatrix}
\end{equation}
and we finally find that
\begin{equation}
[M_i, X^{(i)}X^{(i+1)}]=\hat{X}^{(i)}X^{(i+1)}-X^{(i)}\hat{X}^{(i+1)}
\end{equation}

The relation equivalent to this one in the matrix product formalism (i.e. after making multiple tensor products) can be related to the one found in \cite{PEM} and used to derive the matrix Ansatz for the multispecies periodic ASEP.

Putting this relation back in $U_\mu$ or $T_\mu$ and summing over $i$ cancels out all the terms except for those containing $\hat{X}^{(1)}$ and $\hat{X}^{(L)}$, because all the other $\hat{X}^{(i)}$ appear exactly twice (once in $[M_i,U_\mu]$ and once in $[M_{i+1},U_\mu]$) with opposite signs. Ultimately, we get:
 \begin{align}   \label{MUMThat}
&&\Bigl[\sum_{i=1}^{L-1}M_{i},U_\mu\Bigr]&=\frac{1}{Z_L}  \langle W| A_\mu \hat{X}^{(1)}\prod_{i=2}^{L}X^{(i)} | V \rangle-\frac{1}{Z_L}  \langle W| A_\mu \prod_{i=1}^{L-1}X^{(i)}~\hat{X}^{(L)}| V \rangle\\
&&\Bigl[\sum_{i=1}^{L-1}M_{i},T_\mu\Bigr]&= \langle  \tilde{W}| A_\mu \hat{X}^{(1)}\prod_{i=2}^{L}X^{(i)} |  \tilde{V} \rangle-\langle  \tilde{W}| A_\mu \prod_{i=1}^{L-1}X^{(i)}~\hat{X}^{(L)} |  \tilde{V} \rangle
\end{align}
which we may write as:
 \begin{align}   \label{MUMThat2}
\Bigl[\sum_{i=1}^{L-1}M_{i},U_\mu\Bigr]&=\hat{U}_\mu^{(1)}-\hat{U}_\mu^{(L)}\\
\Bigl[\sum_{i=1}^{L-1}M_{i},T_\mu\Bigr]&=\hat{T}_\mu^{(1)}-\hat{T}_\mu^{(L)} .
\end{align}
so that:
\begin{equation}\label{MUThat}
\Bigl[\sum_{i=1}^{L-1}M_{i},U_\mu T_\mu\Bigr]=\hat{U}_\mu^{(1)}T_\mu-\hat{U}_\mu^{(L)}T_\mu+U_\mu\hat{T}_\mu^{(1)}-U_\mu\hat{T}_\mu^{(L)}
\end{equation}

We now need to check that $[M_0(\mu),U_\mu T_\mu]=-\hat{U}_\mu^{(1)}T_\mu-U_\mu\hat{T}_\mu^{(1)}$ and $[M_L,U_\mu T_\mu]=\hat{U}_\mu^{(L)}T_\mu+U_\mu\hat{T}_\mu^{(L)}$. As before, $M_0$ acts only on site $1$, so that only $X^{(1)}$ is affected, and the same goes for $M_L$ and site $L$.

Let us recall:
\begin{equation}\label{MMu1L}
M_0(\mu)=\begin{bmatrix} -\alpha & \gamma{\rm e}^{-\mu} \\ \alpha{\rm e}^{\mu} & -\gamma  \end{bmatrix}~,~M_L=\begin{bmatrix} -\delta & \beta \\  \delta & -\beta  \end{bmatrix}
\end{equation} 

We calculate:

\begin{equation}\label{MMu1L}
[M_0(\mu),X^{(1)}]=\begin{bmatrix} \gamma{\rm e}^{-\mu}d-\alpha{\rm e}^{\mu}e & (\gamma-\alpha)e \\ (\alpha-\gamma)d & \alpha{\rm e}^{\mu}e- \gamma{\rm e}^{-\mu}d \end{bmatrix}~,~[M_L,X^{(L)}]=\begin{bmatrix} \beta d-\delta e & (\beta-\delta)e \\ (\delta-\beta)d & \delta e- \beta d \end{bmatrix}
\end{equation} 

By projecting these equations on, respectively, $\langle  \tilde{W}| A_\mu$ and $|\tilde{V}\rangle$, we get:
\begin{align}\label{MX1L}
\Bigl[M_0(\mu),\langle  \tilde{W}| A_\mu X^{(1)}\Bigr]&=\langle\tilde{W}|\begin{bmatrix} (\gamma d-\alpha e)A_\mu & A_\mu(\gamma-\alpha)e \\ A_\mu(\alpha-\gamma)d & (\alpha e- \gamma d)A_\mu \end{bmatrix} \nonumber\\
\Bigl[M_L, A_\mu X^{(L)}|\tilde{V}\rangle\Bigr]&=\begin{bmatrix} \beta d-\delta e & (\beta-\delta)e \\ (\delta-\beta)d & \delta e- \beta d \end{bmatrix}|\tilde{V}\rangle
\end{align}
(where we used the second and third relations in (\ref{dehpAlgebra}) to get rid of the $\mu$s). We naturally find the same expressions for $\langle W|$ and $|V\rangle$.

We can then use relations (\ref{VWT}) and (\ref{VW}) to simplify those four equations. We obtain:
\begin{align}\label{MX1L2}
\Bigl[M_0(\mu),\langle  W| A_\mu X^{(1)}\Bigr]&=(\alpha-\gamma)\langle W|A_\mu\begin{bmatrix} 1 & -e \\ d & -1 \end{bmatrix}+(1-q)\langle W|A_\mu \begin{bmatrix} -1 & 0 \\ 0 & 1 \end{bmatrix}\nonumber\\
\Bigl[M_0(\mu),\langle  \tilde{W}| A_\mu X^{(1)}\Bigr]&=(\alpha-\gamma)\langle\tilde{W}|A_\mu\begin{bmatrix} -1 & -e \\ d & 1 \end{bmatrix} \nonumber\\
\Bigl[M_L, A_\mu X^{(L)}|V\rangle\Bigr]&=(\beta-\delta)\begin{bmatrix} -1 & e \\ -d & 1 \end{bmatrix}|V\rangle+(1-q)\begin{bmatrix} 1 & 0 \\ 0 & -1 \end{bmatrix}|V\rangle\nonumber\\
\Bigl[M_L, A_\mu X^{(L)}|\tilde{V}\rangle\Bigr]&=(\beta-\delta)\begin{bmatrix} 1 & e \\ -d & -1 \end{bmatrix}|\tilde{V}\rangle
\end{align}
so that, on site $1$:
\begin{align}
\Bigl[M_0(\mu),\Bigl(\langle  W| A_\mu X^{(1)}\Bigr)\centerdot\Bigl(\langle  \tilde{W}| A_\mu X^{(1)}\Bigr)\Bigr]=&\Bigl((\alpha-\gamma)\langle  W| A_\mu\begin{bmatrix} 1 & -e \\ d & -1 \end{bmatrix}\Bigr)\centerdot\Bigl(\langle  \tilde{W}| A_\mu X^{(1)}\Bigr)\nonumber\\
&+\Bigl((1-q)\langle W|A_\mu \begin{bmatrix} -1 & 0 \\ 0 & 1 \end{bmatrix}\Bigr)\centerdot\Bigl(\langle  \tilde{W}| A_\mu X^{(1)}\Bigr)\nonumber\\
&+\Bigl(\langle  W| A_\mu X^{(1)}\Bigr)\centerdot\Bigl((\alpha-\gamma)\langle\tilde{W}|A_\mu\begin{bmatrix} -1 & -e \\ d & 1 \end{bmatrix}\Bigr)
\end{align}

Seeing that $\begin{bmatrix} 1 & -e \\ d & -1 \end{bmatrix}=X\centerdot \begin{bmatrix} 1 & 0 \\ 0 & -1 \end{bmatrix}$ and that $\begin{bmatrix} -1 & -e \\ d & 1 \end{bmatrix}=\begin{bmatrix} -1 & 0 \\ 0 & 1 \end{bmatrix}\centerdot X$, the first and third part of the right side of this equation cancel out.

What's more, we can write $(1-q) \begin{bmatrix} -1 & 0 \\ 0 & 1 \end{bmatrix}=\frac{(1-q)}{2} \begin{bmatrix} -1 & 0 \\ 0 & 1 \end{bmatrix}\centerdot X+\frac{(1-q)}{2}X\centerdot \begin{bmatrix} -1 & 0 \\ 0 & 1 \end{bmatrix}$ so that
\begin{equation}
(1-q) \begin{bmatrix} -1 & 0 \\ 0 & 1 \end{bmatrix}\centerdot X^{(1)}=- \hat{X}^{(1)}\centerdot  X^{(1)}- X^{(1)}\centerdot \hat{X}^{(1)}
\end{equation}
which means precisely that $[M_0(\mu),U_\mu T_\mu]=-\hat{U}_\mu^{(1)}T_\mu-U_\mu\hat{T}_\mu^{(1)}$. The exact same calculations on the other boundary lead to $[M_L,U_\mu T_\mu]=\hat{U}_\mu^{(L)}T_\mu+U_\mu\hat{T}_\mu^{(L)}$, and this concludes the proof.

\section{Derivation of equation (\ref{MUT}) for the periodic case}
\label{CommutPer}

The periodic case is much simpler than the open one. Equation (\ref{MUT}) takes the form:
\begin{equation}
[M_\mu,T^{per}_\mu]=0
\end{equation}
with
\begin{equation}
T^{per}_\mu={\rm Tr}\Bigl[A_\mu \prod_{i=1}^{L}X^{(i)}\Bigr].
\end{equation}

By the same calculations as in the previous section, we find:
\begin{equation}
\Bigl[\sum_{i=1}^{L-1}M_{i},T^{per}_\mu\Bigr]=(\hat{T}^{per}_\mu)^{(1)}-(\hat{T}^{per}_\mu)^{(L)}
\end{equation}
and we need to check that $[M_{L}(\mu), X^{(L)}A_\mu X^{(1)}]=\hat{X}^{(L)}A_\mu X^{(1)}-X^{(L)}A_\mu\hat{X}^{(1)}$.

We have:
\begin{equation}\label{XXandMper}
X^{(i)}A_\mu X^{(i+1)}=\begin{bmatrix} A_\mu & A_\mu e & e A_\mu & e A_\mu e \\ A_\mu d & A_\mu & e A_\mu d & e A_\mu \\ d A_\mu & d A_\mu e & A_\mu & A_\mu e \\ d A_\mu d & d A_\mu & A_\mu d & A_\mu \end{bmatrix}~~,~~M_{L}(\mu)=\begin{bmatrix} 0 & 0 & 0 & 0 \\ 0 & -q & {\rm e}^{\mu} & 0 \\ 0 & q{\rm e}^{-\mu} & -1 & 0 \\ 0 & 0 & 0 & 0 \end{bmatrix}
\end{equation}
so that:
\begin{align}\label{MXXper}
[M_{L}(\mu), X^{(L)}A_\mu X^{(1)}]&=\begin{bmatrix} 0 & 0 & 0 & 0 \\ {\rm e}^{\mu}d A_\mu-q~A_\mu d & {\rm e}^{\mu}d A_\mu e-q~A_\mu & {\rm e}^{\mu} A_\mu-q~e A_\mu d & {\rm e}^{\mu}A_\mu e-q~e A_\mu \\ {\rm e}^{-\mu}q~A_\mu d-d A_\mu & {\rm e}^{-\mu}q~A_\mu-d A_\mu e & {\rm e}^{-\mu}q~e A_\mu d-A_\mu & {\rm e}^{-\mu}q~e A_\mu-A_\mu e\\ 0 & 0 & 0 & 0 \end{bmatrix}\nonumber\\
&~~~~-\begin{bmatrix} 0 & q({\rm e}^{-\mu}e A_\mu-A_\mu e) & {\rm e}^{\mu}A_\mu e-e A_\mu & 0 \\ 0 & q({\rm e}^{-\mu}e A_\mu d-A_\mu) & {\rm e}^{\mu}A_\mu-e A_\mu d & 0 \\ 0 & q({\rm e}^{-\mu}A_\mu-d A_\mu e) & {\rm e}^{\mu}d A_\mu e-A_\mu & 0 \\  0 & q({\rm e}^{-\mu}A_\mu d-d A_\mu) & {\rm e}^{\mu}d A_\mu-A_\mu d & 0 \end{bmatrix}\nonumber\\
&=\begin{bmatrix} 0 & 0 & 0 & 0 \\ (1-q)A_\mu d & A_\mu(de-q~ed) & (1-q)e A_\mu d & (1-q)e A_\mu\\  (q-1)d A_\mu & (q-1)d A_\mu e & A_\mu(q~ed-de) & (q-1)A_\mu e \\ 0 & 0 & 0 & 0 \end{bmatrix}\nonumber\\
&=(1-q)\begin{bmatrix} 0 & 0 & 0 & 0 \\ A_\mu d & A_\mu & e A_\mu d & e A_\mu \\  -d A_\mu & -d A_\mu e & -A_\mu & -A_\mu e \\ 0 & 0 & 0 & 0 \end{bmatrix}\nonumber\\
&=\hat{X}^{(L)}A_\mu X^{(1)}-X^{(L)}A_\mu\hat{X}^{(1)}
\end{align}
(we used the second and third relation from (\ref{dehpAlgebra}) to get from line one to line two, and the first to get from line two to line three).

Let us note that this same calculation can in fact be used anywhere in the bulk of the open system in order to validate the Ansatz using equs. (\ref{Umui}).

\section{spin-$\frac{1}{2}$ XXZ chain with nondiagonal boundaries}
\label{XXZ}

In this section, we explain how our construction for the open ASEP can be translated for the spin-$\frac{1}{2}$ XXZ chain with non-diagonal boundary conditions \cite{Sandow} and hint at a possible relation with the recent solution of the XXZ chain with a Lindblad boundary drive \cite{SchutzXXZ}.

Let us first define the bulk Hamiltonian of the XXZ spin chain of length $L$:
\begin{equation}\label{HXXZ}
H_{b}=\frac{1}{2}\sum\limits_{k=1}^{L-1}h_i
\end{equation}
with $h_i$ acting as:
\begin{equation}\label{HXXZ}
h_i=\begin{bmatrix} \Delta & 0 & 0 & 0 \\ 0 & -\Delta & 1 & 0 \\ 0 & 1& -\Delta & 0 \\ 0 & 0 & 0 & \Delta \end{bmatrix}
\end{equation}
on sites $i$ and $i+1$ (in the same basis as for equ. (\ref{MMu})), and as the identity on the rest of the chain.

Let us then consider the von Neumann equation for the density operator $\rho$ with the XXZ Hamiltonian with boundary terms $h^0$ and $h^L$ acting only on sites $0$ and $L$:
\begin{equation}\label{vNequ}
\imath\hbar\frac{\partial\rho}{\partial t}=[H,\rho]~~{\rm with}~~H=h_0+H_{b}+h_L
\end{equation}

We will now show how our matrix product construction can be used to find a solution of the stationary equation $\frac{\partial\rho}{\partial t}=0$.

~~

Let us write the deformed Markov matrix $M_{\{\!\mu_i\!\}}$ for the special choice of weights defined by:
\begin{equation}
\{\mu_0=\frac{1}{2}\log{\biggl(\frac{\gamma}{\alpha}\biggr)+\imath\theta_0},~~\mu_i=\frac{1}{2}\log{(q)},~~\mu_L=\frac{1}{2}\log{\biggl(\frac{\delta}{\beta}\biggr)+\imath\theta_L} \}
\end{equation}
which is on the line $\mu=\frac{1}{2} \log{\biggl(\frac{\gamma \delta}{\alpha\beta}q^{L-1}\biggr)}+\imath\mathbb{R}$ and for which $M_{\{\!\mu_i\!\}}$ is Hermitian. The deformed local matrices become:
\begin{equation}\label{MMutheta}
M_0(\mu_0)=\begin{bmatrix} -\alpha & \sqrt{\alpha\gamma}~{\rm e}^{-\imath\theta_0} \\ \sqrt{\alpha\gamma}~{\rm e}^{\imath\theta_0} & -\gamma  \end{bmatrix}~,~ M_{i}(\mu_i)=\begin{bmatrix} 0 & 0 & 0 & 0 \\ 0 & -q &\sqrt{q} & 0 \\ 0 & \sqrt{q}& -1 & 0 \\ 0 & 0 & 0 & 0 \end{bmatrix}~,~M_L(\mu_l)=\begin{bmatrix} -\delta & \sqrt{\beta\delta}{\rm e}^{\imath\theta_L} \\   \sqrt{\beta\delta}{\rm e}^{-\imath\theta_L} & -\beta  \end{bmatrix}
\end{equation}

It is straightforward to check that in this case, we have $M_{\{\!\mu_i\!\}}=\sqrt{q}H+\epsilon$, where $\epsilon$ is a constant, with the boundary matrices being equal to:
\begin{equation}\label{MequalH}
h_0=\frac{1}{2\sqrt{q}}\begin{bmatrix}(1-q-\alpha+\gamma) & 2\sqrt{\alpha\gamma}~{\rm e}^{-\imath\theta_0} \\ 2\sqrt{\alpha\gamma}~{\rm e}^{\imath\theta_0} & (-1+q+\alpha-\gamma) \end{bmatrix}~,~h_L=\frac{1}{2\sqrt{q}}\begin{bmatrix}(-1+q+\beta-\delta) & 2\sqrt{\beta\delta}{\rm e}^{\imath\theta_L} \\ 2\sqrt{\beta\delta}{\rm e}^{-\imath\theta_L} & (1-q-\beta+\delta)  \end{bmatrix}.
\end{equation}

The transfer matrix $U_{\{\!\mu_i\!\}}T_{\{\!\mu_i\!\}}$ defined in (\ref{Umui}) is therefore a solution to equ. (\ref{vNequ}). It might not be a suitable density matrix, as its eigenvalues might not be positive, but we may in any case define one by:
\begin{equation}\label{rhovN}
\rho=\frac{UT\bigl(UT\bigr)^\dag}{Tr\Bigl[UT\bigl(UT\bigr)^\dag\Bigr]}
\end{equation}
(where we omitted to write the dependence in ${\{\!\mu_i\!\}}$).

We can also rewrite expressions (\ref{Umui}) in a way better suited to this situation:
 \begin{align}
U({\mathcal C},{\mathcal C'})&=\frac{1}{Z_L}  \langle \phi| \prod_{i=1}^{L}Y_{\tau_i,\tau_i'} | \psi \rangle\\
T({\mathcal C},{\mathcal C'})&=  \langle  \tilde{\phi}|  \prod_{i=1}^{L}Y_{\tau_i,\tau_i'} |  \tilde{\psi} \rangle .
\end{align}
with
\begin{equation}\label{D0DL}
Y=\begin{bmatrix} N & S_- \\ S_+ & N  \end{bmatrix}
\end{equation}
and
 \begin{align}
N&=A_{\mu_i}\\
S_+&=A_{\frac{\mu_i}{2}}e A_{\frac{\mu_i}{2}}\\
S_-&=A_{\frac{\mu_i}{2}}d A_{\frac{\mu_i}{2}}\\
\langle \phi|&=\langle W|A_{(\mu_0-\frac{\mu_i}{2})}\\
 | \psi \rangle&=A_{(\mu_L-\frac{\mu_i}{2})}| V\rangle\\
\langle \tilde{\phi}|&=\langle  \tilde{W}|A_{(\mu_0-\frac{\mu_i}{2})}\\
 | \tilde{ \psi}\rangle&=A_{(\mu_L-\frac{\mu_i}{2})} | \tilde{V} \rangle.
\end{align}

Matrices $N$, $S_+$ and $S_-$ satisfy a special parametrisation of the $U_q[SU(2)]$ algebra \cite{QGroups}, which is the bulk symmetry of the XXZ chain:
 \begin{align}
[S_+,S_-]&=(\frac{1}{\sqrt{q}}-\sqrt{q})N^2\\
S_+ N&=\frac{1}{\sqrt{q}}N S_+\\
N S_-&=\frac{1}{\sqrt{q}}S_- N
\end{align}

It was surprising to find that this solution has a structure almost identical to that of the Lindblad master equation found in \cite{SchutzXXZ}, where $Y$ is noted $\Omega$ and $\hat{Y}$ is noted $\Xi$. In that case, it seems that $\rho$ can be written in the simpler form $\rho=\frac{UU^\dag}{Tr[UU^\dag]}$, where the algebraic relations satisfied by the boundary vectors $ \langle \phi|$ and $ | \psi \rangle$ might be different from ours. It would be interesting to understand the precise relation between those two {\em a priori} very different situations.


\newpage

 \begin{thebibliography}{99}

\bibitem{MartinReview}  R.~A.~Blythe and  M. R. Evans, 2007,
  {\em Nonequilibrium steady states of matrix-product form: a solver's guide},
 J. Phys. A: Math. Theor. {\bf 40}, R333.

\bibitem{Touchette} H. Touchette, 2009, 
{\em The large deviation approach to statistical mechanics},
 Phys. Rep. {\bf 478} 1.

\bibitem{DerrReview}  B. Derrida, 2007, 
{\em   Non-equilibrium steady  states: fluctuations and large deviations
 of the density and of the current}, J. Stat. Mech.: Theor. Exp.  P07023. Phys. Rev. Lett.  {\bf 87}, 040601 (2001); 

\bibitem{Zia}    B.~Schmittmann and R.~K.~P. Zia, 1995,
 {\em  Statistical mechanics  of driven diffusive systems}, 
  in {\em Phase Transitions and Critical Phenomena vol 17.}, C. Domb and
 J.~L.~Lebowitz Ed., (San Diego, Academic Press).

\bibitem{Schutzrev}    G.~M.~Sch\"utz,
 in {\em Phase Transitions and Critical Phenomena vol 19.}, C. Domb and
 J.~L.~Lebowitz Ed., (Academic Press, San Diego,2001).

\bibitem{Spohn}  H. Spohn, 1991,
{\em Large scale dynamics of interacting particles},
 (Springer-Verlag, New-York).

 \bibitem{Bodineau}  T.~Bodineau,  B.~Derrida, 2004, {\em Current 
    fluctuations in  nonequilibrium diffusive systems: An additivity
  principle},   Phys. Rev. Lett.  {\bf 92}, 180601.

 \bibitem{Bodineau1}  T.~Bodineau,  B.~Derrida,  2005, 
{\em Distribution of currents in  non-equilibrium diffusive systems
 and phase transitions},  Phys. Rev. E {\bf 72}, 066110.

\bibitem{DLSpeer}  B. Derrida, J. L. Lebowitz and E. R. Speer, 2002
{\em Exact free energy functional for a driven diffusive open stationary
 nonequilibrium system},  
 Phys. Rev. Lett.  {\bf 89}, 030601.

\bibitem{Takeuchi} K. A. Takeuchi and M. Sano, {\em Universal Fluctuations of Growing Interfaces: Evidence in Turbulent Liquid Crystals}, Phys. Rev. Lett. {\bf 104},
230601 (2010).

\bibitem{KLS} S.  Katz,  J. L. Lebowitz and H. Spohn, {\em Phase transitions in stationary nonequilibrium states of model lattice systems}, J. Stat. Phys.
 {\bf 34},  497 (1984).

\bibitem{LebSpohn}  J. L. Lebowitz and H. Spohn, {\em A Gallavotti-Cohen Type Symmetry in the Large Deviation Functional for Stochastic Dynamics}, J. Stat. Phys. {\bf 95}, 333 (1999).

\bibitem{Gallavotti} D. J. Evans, E. D. G. Cohen and G. P. Morriss, {\em Probability of second law violations in shearing steady flows},
 Phys. Rev. Lett. {\bf 71}, 2401 (1993). G. Gallavotti and E. D. G. Cohen, {\em Dynamical ensembles in nonequilibrium statistical mechanics},  Phys. Rev. Lett. {\bf 74}, 2694 (1995).

\bibitem{Schutz} G.~M.~Sch\"utz and E. Domany, 1993,
{\em Phase Transitions  in an exactly soluble one-dimensional exclusion
 process}, J. Stat. Phys. {\bf 72}, 277 (1993).

\bibitem{DerridaRep}
 B. Derrida, 1998, 
{\em An exactly soluble non-equilibrium system: the asymmetric simple
 exclusion process}, 
 Phys. Rep.  {\bf 301}, 65.

\bibitem{Varadhan}  S. R. S. Varadhan, 1996, {\em  The complex story of simple
exclusion}, in  It$\hat{\rm o}$ stochastic calculus and probability theory, 385 
(Springer, Tokyo). 

\bibitem{Gorissen} M. Gorissen and C. Vanderzande, {\em Finite size scaling of current fluctuations in the totally asymmetric exclusion process },  J. Phys. A: Math. Theor. {\bf 44} 115005 (2011).

\bibitem{Doucot}  B. Derrida, B. Dou\c{c}ot and P.-E. Roche, 2004,
 {\em Current fluctuations in the one-dimensional symmetric exclusion
 process with open boundaries}, J. Stat. Phys. {\bf 115} 717.

\bibitem{deGierNew}
  J.~de~Gier, F.~H.~L. Essler, 2011,  {\em Current large deviation
 function for the open asymmetric simple exclusion process},
 arXiv:1011.3235. 

  \bibitem{PaulK}  P. L. Krapivsky, S. Redner and E. Ben-Naim, 2010,  {\it  A
Kinetic View of Statistical Physics} (Cambridge: Cambridge University
Press).

\bibitem{Chou}   T. Chou, K. Mallick and R. K. P. Zia, {\em Non-equilibrium statistical mechanics: From a paradigmatic model to biological transport},
 Rep. Progr.  Phys.  {\bf 74},  116601 (2011).

\bibitem{Zia2} D. A. Adams, B. Schmittmann and R. K. P. Zia, {\em Far-from-equilibrium transport with constrained resources}, J. Stat. Mech.: Theor. Exp. P02012 (2009).

\bibitem{SasamSpohn} T. Sasamoto and H. Spohn, {\em The one-dimensional KPZ equation: an exact solution and its universality},  Phys. Rev. Lett. {\bf 104}, 230602 (2010).
  G. Amir, I. Corwin and J. Quastel, {\em Probability Distribution of the Free Energy of the Continuum Directed Random Polymer in 1+1 dimensions}, Comm. Pure Appl. Math.  {\bf 64},
   466 (2011).
  P. L.  Ferrari, {\em From interacting particle systems to random matrices} J. Stat. Mech.: Theor. Exp.  P10016  (2010).

\bibitem{KPZ} M. Kardar, G. Parisi and Y.-C. Zhang, {\em Dynamic scaling of growing interfaces}, Phys. Rev. Lett.
  {\bf 56}, 889 (1986).
  T. Halpin-Healy, Y.-C.~Zhang, {\em Kinetic Roughening, Stochastic Growth, Directed Polymers \& all that}, Phys. Rep.  {\bf 254}, 215 (1995).

 \bibitem{FerrariPatrick}  P. L. Ferrari,  {\em  From Interacting
 particle systems to random matrices}, J. Stat. Mech.: Theor. Exp.  P10016  (2010).

\bibitem{Sasamoto} T. Sasamoto, 2007, 
{\em Fluctuations of the one-dimensional asymmetric exclusion process
using random matrix techniques},  J. Stat. Mech.: Theor. Exp.  P07007.

\bibitem{Johansson} K. Johansson, {\em Shape Fluctuations and Random Matrices}, Comm. Math. Phys.  {\bf 209},  437  (2000)
 T. Kriecherbauer and J. Krug, {\em A pedestrian's view on interacting particle systems, KPZ universality, and random matrices }, J. Phys. A: Math.  Theor. {\bf 43}, 403001 (2010).
 I. Corwin, {\em The Kardar-Parisi-Zhang equation and universality class},  Random Matrices, Theory and Appl.    {\bf 1},  (2012). 

\bibitem{TracyWidom} C.A.  Tracy, H.  Widom,  {\em  Total Current Fluctuations in ASEP},  J. Math. Phys. {\bf 50}, 095204 (2009). 

\bibitem{Corteel} S. Corteel and L. Williams, {\em Tableaux combinatorics for the asymmetric exclusion process and Askey-Wilson polynomials}, Duke Math. J. {\bf 159}, 385-415 (2011).

\bibitem{Viennot} X.~G.~Viennot, {\em Canopy of binary trees, Catalan tableaux and the asymmetric exclusion process}, FPSAC 2007, Formal Power Series and Algebraic Combinatorics.

\bibitem{Krug}  J. Krug, {\em Boundary-induced phase transitions in driven diffusive systems},  Phys. Rev. Lett. {\bf 67}, 1882 (1991).

\bibitem{DeDoMuk}  B. Derrida, E. Domany and D. Mukamel, 1992
{\em  An exact solution of a one-dimensional asymmetric exclusion
 model with open boundaries},  J. Stat. Phys. {\bf 69},  667. 

\bibitem{DEHP}
 B. Derrida, M.~R.~Evans, V. Hakim, V. Pasquier, 1993,
 {\em Exact solution of a 1D  asymmetric exclusion
model  using a matrix formulation}, 
J. Phys. A: Math. Gen. {\bf 26}, 1493.

 \bibitem{DEMal}  B. Derrida, M. R. Evans and
 K. Mallick, 1995, {\em Exact Diffusion Constant 
 of a  One-Dimensional  Asymmetric Exclusion Model with Open Boundaries},
 J. Stat. Phys. {\bf 79}, 833.

\bibitem{DLeb}
 B. Derrida, J. L. Lebowitz, 1998,
{\em Exact  large deviation function in the asymmetric exclusion process},  
 Phys. Rev. Lett.  {\bf 80}, 209.

\bibitem{Sylvain4} S. Prolhac, 2010, 
{\em Tree structures for the current fluctuations in the exclusion
 process},  J. Phys. A: Math. Theor. \textbf{43}, 105002.

\bibitem{PEM}
S. Prolhac, M. R. Evans, K. Mallick, 2009,
{\em The matrix product solution of the multispecies
partially asymmetric exclusion process},
J. Phys. A: Math. Theor. \textbf{42}, 165004.

\bibitem{SasaPasep1}  T. Sasamoto, {\em One-dimensional partially asymmetric simple exclusion process with open boundaries: orthogonal polynomials approach},  J. Phys. A: Math. Gen. {\bf 32}, 7109 (1999).
M.  Uchiyama  and M. Wadati, {\em Correlation Function of Asymmetric Simple Exclusion Process with Open Boundaries},
J.Nonlinear Math. Phys. {\bf 12}, 676 (2005).

\bibitem{DSimon} D. Simon, {\em Construction of a Coordinate Bethe Ansatz for the asymmetric simple exclusion process with open boundaries}, J. Stat. Mech.: Theor. Exp. P07017 (2009) .

\bibitem{deGier1}
  J.~de~Gier, F.~H.~L. Essler, 2005,
  {\em  Bethe Ansatz solution of the Asymmetric Exclusion Process
 with Open Boundaries}, Phys. Rev. Lett.  {\bf 95}, 240601.

\bibitem{DLeb2}
 B. Derrida and C. Appert, {\em Universal large deviation function of the Kardar-Parisi-Zhang equation in one dimension},  J. Stat. Phys. {\bf 94}, 1 (1999).
 B. Derrida, M. R. Evans, {\ Bethe Ansatz Solution for a Defect Particle in the Asymmetric Exclusion Process}, J. Phys. A: Math. Gen. {\bf 32}, 4833  (1999)

\bibitem{LazarescuMallick} A. Lazarescu and K. Mallick, {\em An Exact Formula for the Statistics of the Current in the TASEP with Open Boundaries
}, J. Phys. A: Math. Theor. {\bf 44} 315001 (2011).

\bibitem{GLMV} M. Gorissen, A. Lazarescu, K. Mallick and C. Vanderzande, {\em Exact Current Statistics of the ASEP with Open Boundaries
}, Phys. Rev. Lett. 109, 170601 (2012).

\bibitem{Kurchan} C. Giardina,  J. Kurchan and L. Peliti, {\em Direct evaluation of large-deviation functions}, Phys. Rev. Lett.
  {\bf 96} 120603 (2006).  C. Giardina,  J. Kurchan,  V. Lecomte  and  J. Tailleur, {\em Simulating rare events in dynamical processes},
 J. Stat. Phys.  {\bf 145}, 787  (2011).

\bibitem{Bertini}  L. Bertini, A. De Sole, D. Gabrielli,
  G. Jona-Lasinio and C. Landim,
 {\em Macroscopic current fluctuations in stochastic lattice gases},  Phys. Rev. Lett.  {\bf 94}, 030601 (2005). 

\bibitem{AppDerr} C. Appert-Rolland, B. Derrida, V. Lecomte and F. Van Wijland, {\em Universal cumulants of the current in diffusive systems on a ring}, Phys. Rev. E {\bf 78}, 021122.

\bibitem{SchutzXXZ} D. Karevski, V. Popkov and G.M. Schütz, {\em Exact matrix product solution for the boundary-driven Lindblad $XXZ$-chain},  	arXiv:1211.7010

\bibitem{DiscTime} N. Rajewsky, L. Santen, A. Schadschneider and M. Schreckenberg, {\em The asymmetric exclusion process: Comparison of update procedures}, J. Stat. Phys. {\bf 92}, 1-2 (1998).

\bibitem{Sandow} S. Sandow, {\em Partially Asymmetric Exclusion Process with Open Boundaries},  Phys. Rev. E {\bf 50}, 2660.

\bibitem{DonskVar} J. D. Deuschel and D. W. Stroock, {\em Large Deviations}, Academic Press, Boston (1989).

\bibitem{SEnsemble} R. L. Jack and P. Sollich, {\em Large deviations and ensembles of trajectories in stochastic models}, Prog. Theoret. Phys. Suppl., 184:304–317 (2010).

\bibitem{EvansTraffic} M. R. Evans, N. Rajewsky and E. R. Speer, {\em Exact solution of a cellular automaton for traffic}, J. Stat. Phys.  {\bf 95}, 1-2  (1999).

\bibitem{Stinch} M. Depken and R. Stinchcombe, {\em Exact joint density-current probability function for the asymmetric exclusion process}, Phys. Rev. Lett. {\bf 93}, 040602 (2004).

\bibitem{QGroups} M. Chaichian and A. Demichev, {\em Introduction to Quantum Groups}, World Scientific Pub. (1996)






\end{thebibliography}
\end{document}